\documentstyle[psfig]{article}
\newcommand{\ket}[1]{ | \, #1  \rangle}
\newcommand{\one}{\mbox{$1 \hspace{-1.0mm}  {\bf l}$}}
\begin{document}
\title{\bf Quantum entanglement and classical communication through a
depolarising channel }
\author{Dagmar Bru\ss,\\
{\small\em Institut f\"ur Theoretische Physik, Universit\"at Hannover,
Appelstr. 2, D-30167 Hannover, Germany}\\
Lara Faoro, \\
{\small \em Dipartimento di Fisica, Politecnico di Torino, Via Duca degli
Abruzzi 24, I-10129 Torino, Italy}\\
Chiara Macchiavello, \\
{\small \em Dipartimento di Fisica ``A.Volta" and INFM-Unit\`a di Pavia, 
Via Bassi 6, I-27100 Pavia, Italy}\\
G.Massimo Palma\\
{\small \em Dipartimento di Scienze Fisiche ed Astronomiche and INFM-Unit\`a 
di Palermo,Via Archirafi 36, I-90123 Palermo, Italy}}
\date{\today}
\maketitle
\begin{abstract}
We analyse the role of entanglement for transmission of classical
information through a memoryless depolarising channel. Using the isotropic
character of
this channel we prove analytically that the mutual information cannot be 
increased by encoding classical bits into entangled states of two qubits.
\end{abstract}

\section{Introduction}
Entanglement is probably the most important resource in quantum information
processing:
quantum teleportation \cite{teleport1}, quantum cryptography \cite{qcrypto}, 
quantum computation \cite{qcomp}, quantum error correction
\cite{errorcorr1} are only some examples of its ubiquitous crucial role. 
In all the instances mentioned above its
main use is transmitting, processing, 
correcting quantum information. In this paper we would like to analyse 
whether it is useful in one particular 
scenario of transmission of classical information along a noisy 
channel: the depolarising channel.  
It has been proved that when multiple uses of the  channel are allowed
entangled states can maximise 
distinguishability in a particular case of noisy channel: the two-Pauli
channel  \cite{disting}. 
It is however common belief that, due to their fragility in the presence of
noise, entangled states cannot 
improve the transmission of classical information when used as signal
states,  at least when the noise is 
isotropic - in the sense that will be explained in the following sections -
as in the case of the depolarising channel. 
Although there is numerical evidence in this direction \cite{disting}
no explicit proof has
been produced so far, at least at the 
best of the authors' knowledge. 
The scope of this paper is precisely to
prove that the mutual information 
cannot be increased when signals of entangled states 
of two qubits are used with a
memoryless depolarising channel.

\section{Description of the Channel}

In the following we will consider a memoryless quantum  channel acting on
individual qubits. The channel   is described by ``action" operators 
\cite{kraus} $A_k$ satisfying $\sum_k A^{\dagger}_kA_k = \one$
such that if we send through the channel a qubit in a state described by
the density operator
$\pi$ the output qubit state is given by the map

\begin{equation}
\pi \longrightarrow \Phi (\pi ) =\sum_k A_k \pi A_k^{\dagger}
\end{equation}

By definition a channel is memoryless when its action on arbitrary 
signals
 $\pi_i$, consisting  of $n$ qubits (including entangled ones), is given by

\begin{equation}
\Phi (\pi_i ) = \sum_{k_1\cdots k_n} 
(A_{k_n}\otimes \cdots \otimes
A_{k_1}) \pi_i (A^{\dagger}_{k_1}\otimes \cdots \otimes
A^{\dagger}_{k_n})
\end{equation}

For the depolarising channel, on which we will concentrate our attention,
the action operators take the following simple form 

\begin{equation}
A_0 = \frac{1}{2}\sqrt{1+3\eta}\, \one , \hspace{3cm} 
A_{x,y,z} = \frac{1}{2}\sqrt{1-\eta}\, \sigma_{x,y,z}\ .
\end{equation}
Here $\sigma_{x,y,z}$ are the Pauli matrices, which are transformed under the 
action of the channel as

\begin{equation}
\sum_k A_k \sigma_l A^{\dagger}_k = \eta\sigma_l\;.
\label{shrink}
\end{equation}

As we can see, the depolarising channel can be specified by the parameter
$\eta$, whose meaning will be clear in the following section, or
equivalently by the error probability $p_e=3(1-\eta)/4$. 
This gives us a complete description of the channel for any
input state composed of an arbitrary number of qubits.

\section{Entanglement and mutual information}

In the simplest scenario the transmitter can send one qubit at a time along
the channel.
In  this case the codewords will be restricted to be the tensor product of
the states of
the individual qubits. Quantum mechanics however allows also the
possibility to entangle
multiple uses of the channel. For this more general strategy it has been
shown that the 
amount of reliable information which can be transmitted per use of the
channel is 
given by \cite{Westmoreland,Holevo}

\begin{equation}
C_n =\frac{1}{n}{\mbox{sup}}_{\cal E} I_n(\cal E)\ \ ,
\end{equation}

where ${\cal E}=\{p_i,\pi_i\}$  with $p_i\geq 0, \sum p_i=1$ 
is the input ensemble of states $\pi_i$ of  $n$ -- generally entangled 
-- qubits
and $I_n({\cal E})$ is the mutual information

\begin{equation}
I_n({\cal E}) = S(\rho ) - \sum_i p_i S(\rho_i)\ \ ,
\label{def-I}
\end{equation}

where the index $n$ stands for the number of uses of the channel. Here

\begin{equation}
S(\chi ) = - {\mbox{tr}}(\chi \log \chi )
\end{equation}

is the von Neumann entropy,  $\rho_i = \Phi (\pi_i)$ are 
the density matrices of the outputs 
and $\rho = \sum_i p_i \rho_i$. 
Logarithms are taken to base 2.
The advantage of the expression 
(\ref{def-I}) is that it includes an optimisation over 
all possible POVMs at the output, including collective ones.
Therefore no explicit maximisation procedure for the decoding at the
output of the channel is needed.

The interest for the possibility of using entangled states as channel input
is that it cannot generally be excluded that
$I_n({\cal E})$ is  superadditive for entangled
inputs, i.e.
we might have $I_{n+m}> I_n + I_m$ and therefore $C_n > C_1$.

In this scenario the classical capacity $C$ of the channel is defined as

\begin{equation}
C=\lim_{n\to \infty}  C_n\;.
\label{capac}
\end{equation}

For the depolarising channel a lower bound on $C$ is given by the one-shot 
capacity $C_1$ (see \cite{Westmoreland}), 
while upper bounds are given in \cite{bds}. 
In this paper we will not attempt to find the classical capacity of the
depolarising
channel, as this would imply analysing signals with any 
degree of entanglement between $n$ qubits and performing 
the limit $n\to \infty$. 
%This goes beyond present knowledge. 
We will restrict ourselves to the simplest non-trivial case,
namely $n=2$, and we will find the maximal mutual information $I_2({\cal E})$.

The question we want to address is: is it possible to increase the mutual
information 
%$I_2({\cal E})$  
by entangling the two qubits, i.e. is $C_2 > C_1$ for
the depolarising
channel? To answer this question with ``no''
we will show that  due to the isotropy 
of the depolarising channel  the mutual information 
for orthogonal entangled states of two qubits 
 depends only on  the degree of entanglement, and in
particular that it is a decreasing function of the entanglement.
We anticipate that, as we will show later, the mutual information cannot be
increased 
by an alphabet of non-orthogonal states.

To simplify our analysis we express the signal states in  the ``Bloch
vector" representation \cite{Mahler}

\begin{equation}
\pi = \frac{1}{4} \{ \one\otimes\one 
+ \one\otimes \sum_k\lambda_{k}^{(2)} \sigma_{k} +
\sum_k\lambda_{k}^{(1)}
\sigma_{k} \otimes\one
+\sum_{kl} \chi_{kl}\sigma_{k}\otimes\sigma_{l} \}\ \ ,
\end{equation}

where as before $\sigma_k$ are the Pauli operators of the two qubits. 
Using Eq. (\ref{shrink}) the output of the channel can be written as

\begin{equation}
 \rho = \Phi (\pi) = \frac{1}{4} \{ \one\otimes\one + 
\eta \sum_k(\one\otimes \lambda_{k}^{(2)}\sigma_{k} +  
\lambda_{k}^{(1)} \sigma_{k}
\otimes\one)
+ \eta^2\sum_{kl}\chi_{kl}\sigma_{k}\otimes\sigma_{l}\}\;.
\end{equation}

This shows that the output states are linked to the input signals by an
isotropic
shrink of the Bloch vectors $\lambda_{k}^{(1)} , 
\lambda_{k}^{(2)}$ by a factor $\eta$ 
and of the tensor $\chi_{kl}$ by a factor $\eta^2$.

We will now consider the following general signal state 

\begin{eqnarray}
|\pi_{i}\rangle & = & \cos\vartheta_i |00\rangle_i + \sin\vartheta_i
|11\rangle_i\;.
\label{signals}
\end{eqnarray}

In the above equation $\vartheta_i\in[0,\pi/4]$ parametrises 
the degree of entanglement between the two qubits that carry the signal,
which we will  measure with the usual entropy of entanglement
$S(\tau)=-{\mbox{tr}}\{\tau\log\tau\}$ \cite{entang}, where
$\tau = {\mbox{tr}}_1 \pi_i = {\mbox{tr}}_2 \pi_i$ denotes the reduced 
density operator of one qubit
(note that $\tau$  is the same for both qubits as the trace can be taken
indifferently over any of the two ). For the state (\ref{signals}) we have
$S_i= - \cos^2\vartheta_i \log \cos^2\vartheta_i - \sin^2\vartheta_i\log
\sin^2\vartheta_i$.

We point out that  this is the most general choice for a signal state.
This can be easily verified observing that any 
pure state $\ket{\Psi}$  of two qubits can be decomposed in the Schmidt
basis as 
follows \cite{Schmidt}

\begin{eqnarray}
\ket{\Psi}=c_1\ket{u_1}\ket{v_1}+c_2\ket{u_2}\ket{v_2}\;,
\end{eqnarray}
where $c_i$ are real and $c_1^2+c_2^2=1$, 
while $\{\ket{u_i}\}$ and $\{\ket{v_i}\}$
represent orthogonal bases for the first and the second qubit, respectively. 
These two bases will not  in general  have the same orientation in the Bloch
vector representation, but for
our purpose
this is not an impediment because of the isotropy of the depolarising
channel. In other
words, since there is no privileged direction for the choice of basis
$\{\ket{0},\ket{1}\}$ 
 - they do not even have to be the same for each of the two qubits - 
we are free to let them coincide with the two respective Schmidt bases.

After the action of the depolarising channel, the density 
operator corresponding to the input state (\ref{signals})
takes the form

\begin{eqnarray}
\rho_i =  
\frac{1}{4}\{ \one\otimes\one + \eta\cos 2\vartheta_i (\one\otimes\sigma_z + 
\sigma_z\otimes\one )
+ \eta^2(\sigma_z\otimes \sigma_z
+ \sin 2\vartheta_i ( \sigma_x \otimes\sigma_x  
- \sigma_y\otimes \sigma_y )) \}\;.
\label{output}
\end{eqnarray}

In the basis $\{\ket{00},\ket{01},\ket{10},\ket{11}\}$ 
the output $\rho_i$ reads

\[
\rho_i = \frac{1}{4}\left( \begin{array}{cccc}
1+2\eta\cos 2\vartheta_i  +\eta^2 & 0 & 0 & 2\eta^2\sin 2\vartheta_i \\
0 & 1-\eta^2 & 0 & 0 \\
0 & 0 & 1-\eta^2 & 0 \\
2\eta^2\sin 2\vartheta_i   & 0 & 0 &1-2\eta\cos 2\vartheta_i +\eta^2 
\end{array} \right)
\]

The corresponding eigenvalues are 

\begin{eqnarray}
\alpha_1 = \alpha_2 & = &\frac{1}{4}( 1-\eta^2)\\
\alpha_{3,4}(\vartheta_i)
& = &  \frac{1}{4}(1 + \eta^2 \pm 2\eta\sqrt{\cos^2 2\vartheta_i + 
\eta^2\sin^2 2\vartheta_i})\;. 
\end{eqnarray}
They depend on the degree of entanglement, but 
are independent of the choice of basis.

For maximisation of the second term of the mutual information $I_2$,
given in Eq. (\ref{def-I}), it is 
sufficient to optimise independently  terms of the form

\begin{equation}
S(\rho_i)=\sum_{j=1}^4 \alpha_j(\vartheta_i) \log\alpha_j(\vartheta_i) \;.
\label{mutual}
\end{equation}
Remember that the {\it a priori} probabilities add to one, so neither the 
number of input states nor their probabilities enter in the maximisation
of the second term. As we will show later, the first term in (\ref{def-I})
and therefore the mutual information is maximised for a set of orthogonal
and equally probable states.

We can search the extremum of (\ref{mutual}) 
analytically leading to the requirement
\begin{equation}
\cos2\vartheta_i \sin2\vartheta_i 
\cdot \{\log(1 + \eta^2 +  2\eta\sqrt{\cos^2 2\vartheta_i + \eta^2\sin^2
2\vartheta_i})-
\log(1 + \eta^2 -  2\eta\sqrt{\cos^2 2\vartheta_i + \eta^2\sin^2
2\vartheta_i})\}=0 \ \ .
\end{equation}
For $\eta\neq 0$ there are two solutions, namely
\begin{equation}
\vartheta_i =0 \ ,
\end{equation}
which corresponds to non-entangled input states
and turns out to be the maximum, and
\begin{equation}
\vartheta_i =\pi/4 \ ,
\end{equation}
which means maximal entanglement of the inputs
and corresponds to the minimum of $I_2$.

The explicit expression for the maximal mutual 
information corresponding to four
equiprobable orthogonal non-entangled input states is given by

\begin{eqnarray}
I_2^{max} & = & 
 (1 + \eta)\log (1 + \eta)+ 
 (1- \eta)\log (1- \eta)\;.
\label{maxinfo}
\end{eqnarray}

Notice that this is equivalent to twice the one-shot capacity. 
So we have shown that in this case entanglement does
not help to increase the mutual information \cite{dense}.

As promised, we now justify our choice of orthogonal states by
proving that the mutual information cannot be increased 
by using non-orthogonal alphabets. 
We were able to maximise the two parts of
the mutual information, defined in equation (\ref{def-I}),
independently of each other for the  following reasons.
Note first that $S(\rho)$ is maximal for
$\rho=\frac{1}{4}\one_1\otimes\one_2$.
For the depolarising channel this form is achieved by any set of four
orthogonal equiprobable input states.
The minimum of the term $\sum_i p_i S(\rho_i)$
in the mutual information, however,  does not depend on the 
orthogonality of the states, as only the eigenvalues of the output states 
determine the extremum.
We noticed that each $S(\rho_i)$ can be minimised independently for each
input state $\pi_i$.
As we have shown before, $S(\rho_i)$ is minimal, and reaches the same 
minimum value, for any choice of input states of non-entangled qubits.
Note that due to these reasons the same  mutual information could be also
reached
by using a larger alphabet of non-orthogonal non-entangled states and
adjusted probabilities, but
it  can never be improved beyond the maximum value given in 
(\ref{maxinfo}).

\begin{figure}[hbt]
\vspace*{-1cm}
\centerline{\psfig{file=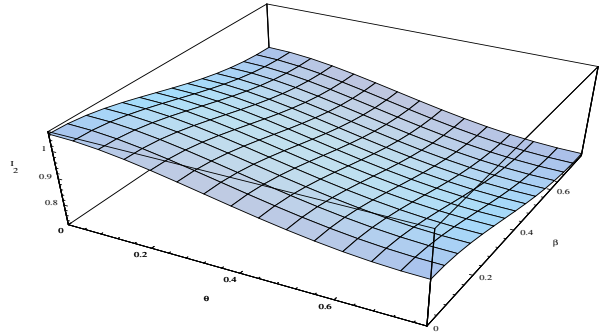,height=8cm,width=8cm}}
\caption{Mutual information for two uses of the depolarising
   channel as a function of $\vartheta$ and
$\beta$ for $\eta=0.8$.}
\label{fig1}\end{figure}

\begin{figure}[hbt]
\centerline{\psfig{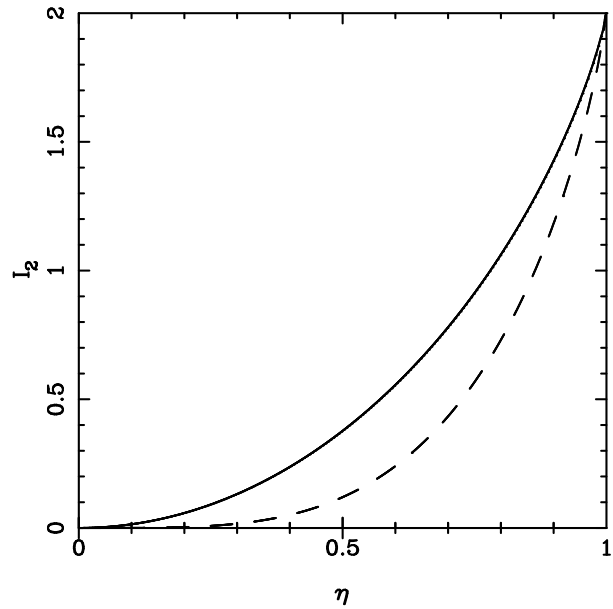}}
\caption{Mutual information 
for two uses of the depolarising channel
as a function of $\eta$ for %$\vartheta=\beta=0$
product input states
(full line) and %$\vartheta=\beta=\pi/4$ 
maximally entangled input states (dashed line).}
\label{fig2}\end{figure}

As an illustration, in figure \ref{fig1}
we show  the mutual information for the following set of equally
probable orthogonal states
\begin{eqnarray}
|\pi_{1}\rangle & = & \cos\vartheta |00\rangle + \sin\vartheta
|11\rangle\nonumber\\
|\pi_{2}\rangle & = & \sin\vartheta |00\rangle -  \cos\vartheta
|11\rangle\nonumber\\
|\pi_{3}\rangle & = & \cos\beta |01\rangle + \sin\beta |10\rangle\nonumber\\
|\pi_{4}\rangle & = & \sin\beta |01\rangle - \cos\beta |10\rangle
\end{eqnarray}
as a function of $\vartheta$ and $\beta$ for 
$\eta=0.8$. As we can see, the mutual 
information is a decreasing function of the degree of entanglement.
In figure \ref{fig2} we report the mutual information as a function of $\eta$
for uncorrelated states and maximally entangled states. 
%(dotted line). 
As proved above, we can see that uncorrelated signals lead to 
a higher mutual information for any channel with $0<\eta <1$.

To summarise, we have asked whether classical communication through a
depolarising channel can be improved by entangling two uses of the channel. 
Our analytical results show that this is not the case:
the mutual information is maximised when using orthogonal equiprobable
non-entangled states. 
The generalisation to more than two qubits
remains an open problem \cite{comment}.

\section{Acknowledgements}
We would like to thank 
C.H. Bennett, C. Fuchs, R. Jozsa,
G. Mahler and J. Schlienz for
helpful discussions.
In particular,  we thank A. Peres for constructive critics.
This work was supported in part by the European TMR Research 
Network ERB 4061PL95-1412 ``The physics of quantum information", by
Ministero dell'Universit\`a e della Ricerca Scientifica e Tecnologica
under the project ``Amplificazione e rivelazione di radiazione quantistica''
and by Deutsche Forschungsgemeinschaft under grant SFB 407.
Part of this work was completed during the 1998 workshops on quantum
information organised by  ISI Foundation - Elsag-Bailey and by the 
Benasque Center for Physics.

\begin{thebibliography}{99}
\bibitem{comment}
Notice that our line of argument is based on the
Schmidt decomposition and therefore it cannot be directly generalised to
more than two uses of the channel.
\end{thebibliography}
\end{document}